\documentclass{article}
\usepackage{spconf,amsmath,graphicx}
\usepackage{enumerate}
\usepackage{diagbox}
\usepackage{tabularx}
\usepackage{xfrac}

\title{A BIN ENCODING TRAINING OF A SPIKING NEURAL NETWORK BASED VOICE ACTIVITY DETECTION }
\begin{document}

\name{Giorgia Dellaferrera$^{\star \dagger}$ \qquad Flavio Martinelli$^{\star \dagger}$ \qquad Milos Cernak$^{\dagger}$}

\address{$^{\star}$ Ecole Polytechnique F\'ed\'erale de Lausanne (EPFL), Switzerland \\
    $^{\dagger}$ Logitech Europe S.A., Lausanne, Switzerland }
%
%
%

%
\maketitle
\begin{abstract}
Advances of deep learning for Artificial Neural Networks (ANNs) have led to significant improvements in the performance of digital signal processing systems implemented on digital chips. Although recent progress in low-power chips is remarkable, neuromorphic chips that run Spiking Neural Networks (SNNs) based applications offer an even lower power consumption, as a consequence of the ensuing sparse spike-based coding scheme.
In this work, we develop a SNN-based Voice Activity Detection (VAD) system that belongs to the building blocks of any audio and speech processing system. We propose to use the bin encoding, a novel method to convert log mel filterbank bins of single-time frames into spike patterns. We integrate the proposed scheme in a bilayer spiking architecture which was evaluated on the QUT-NOISE-TIMIT corpus. Our approach shows that SNNs enable an ultra low-power implementation of a VAD classifier that consumes only $3.8 \mu$W, while achieving state-of-the-art performance.
\end{abstract}
\section{Introduction}
\label{sec:intro}

Recent development of voice user interfaces offers communication between users and consumer electronics devices as natural as possible. 
Enabling hands-free activation and verbal control represent an appealing technology adopted by most of manufacturers and masses of customers. 
Moving computation from the cloud to the edge is characterised by looking for alternative computational platforms that can perform high performance computation but operate on a battery for a longer time (about several months). One promising candidate of those alternative platforms is neuromorphic hardware. For example, recent benchmarking of keyword spotting efficiency~\cite{Blouw} reports more than 5 times less energy consumption of the Intel neuromorphic chip Loihi~\cite{Loihi} comparing to Movidius neural compute stick designed for the edge processing.

An intrinsic property of neuromorphic computing is running Spiking Neural Nets (SNNs) instead of Artificial Neural Nets (ANNs). SNNs are known as the third generation of neural networks. They are inspired by the human brain computation, from which they inherit the sparse and asynchronous nature of the information, leading to the properties of high power efficiency and robustness to noise. These features make SNNs an ideal candidate to build low-power models for speech detection and processing. 
In this paper, we explore SNN-based Voice Activation Detection (VAD). A VAD system belongs to the basic building blocs of any audio and speech processing system, and, as it is always on, it needs to be power efficient. The VAD systems are well studied and recently several very low power algorithms~\cite{Price,Meoni} and the circuits~\cite{Yang} were proposed.

Several approaches designing SNN models for audio processing have already been developed as well. Some algorithms are based on supervised learning techniques, such as the nonrecurrent SNN developed in \cite{TavanaeiM17} that extracts the spike features from speech signals, and the framework for sound recognition presented in \cite{Yu}. Unsupervised learning is exploited in other approaches for both feature extraction \cite{Tavanei_Maida} and isolated word classification utilizing the principles of Self-Organizing Maps \cite{Hazan,Rumbell}. 
Some studies use instead a reservoir based-technique to solve the task of isolated digit recognition \cite{VerstraetenLSM, Verstraeten}. However, to our knowledge, no prior published work explored the application of SNN for VAD. Most of the existing encoding methods (Poisson, time-to-first-spike) have been developed for tasks such as word recognition \cite{ConvSNN}. Speech representation of isolated words is generally characterized by high interclass variability and high intraclass correlation. The \textit{Voice} and \textit{No voice} classes of the VAD task lack these features and therefore the existing encoding methods, designed specifically for multi-time frame speech representation, are not optimal to represent a single time frame desirable for voice activity detection. To overcome these drawbacks, we developed a novel encoding method, the bin encoding, whose strength resides in the possibility of capturing the information on all frequency filters.


\section{Proposed methods}
\label{sec:methods}
\begin{figure}
    \centering
    \includegraphics[width=0.35\textwidth]{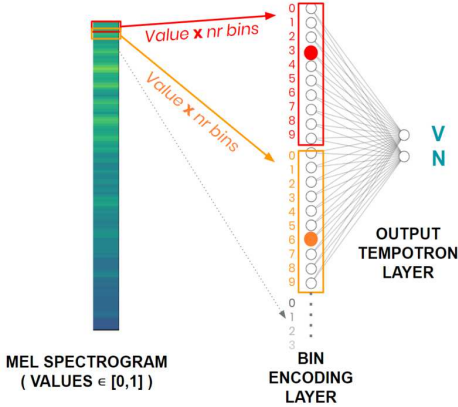}
    \caption{The bin encoding architecture proposed in this work.}
    \label{fig:bin_enc}
\end{figure}

The proposed VAD method consists of a preprocessing stage followed by a bilayer spiking architecture, as illustrated in Fig.\ \ref{fig:bin_enc}.
In the preprocessing stage, the audio signal is extracted into log mel filterbanks, as follows.
First, we used a sample rate of 16kHz, 128 mel frequency bins, 40 ms time frame length, and 50\% overlap between frames, leading to a the classification which is performed every 20 ms. Next, the preprocessing is completed through the normalization of the frequency bins within the interval $[0,1]$ w.r.t.\ the minimum and maximum values of all the frequencies computed over the entire data set.

Our spiking architecture is composed of two fully connected layers, \emph{i.e.}, the input encoding layer and the output layer. The input layer is formed by the one-dimensional arrangement of the groups of $N_{in}$ neurons, each of them associated with one of the 128 frequency bands. We carefully chose $N_{in} = 10$, leading to a total number of input neurons of $N_{in}\times128 = 1280$. Such an architecture represents the basis of the bin encoding.

The ouput layer consists of two neurons, with one of them firing in response to speech and the other to non-speech. We denote these neurons as $V$ and $N$, respectively. For the training of this layer, we relied on the Maximum-Margin Tempotron temporal learning rule devised in~\cite{SNNRobustSoundClassific}. This approach combines the Tempotron learning rule~\cite{tempotron} with the maximum-margin classifier~\cite{Cortes1995}.
It is worth noticing that the Tempotron learning rule is a biologically plausible model of supervised learning suitable for decoding information embedded in spatio-temporal spike patterns. The adopted neuron model is a Leaky-Integrate-and-Fire (LIF) neuron driven by exponentially decaying synaptic currents generated by its afferents. The subthreshold membrane voltage ($V(t)$) is a weighted sum of postsynaptic potentials (PSPs) induced by  the $j$-th incoming spike, taking the form
\begin{equation}
    V(t) = \sum_j \omega_j \sum_{t_j}K(t-t_j)+V_{rest} \label{eq:membrane}
\end{equation}
with $\omega_j$ being a synaptic weight of the incoming $j$-th synapse, $t_j$ the spike time of the presynaptic $j$-th neuron, $V_{rest}$ the resting membrane potential, and $K(t-t_j)$ the kernel function representing the effect of the presynaptic spikes on $V(t)$. This latter quantity is expressed as
    \begin{equation}
        K(t-t_j) = V_0 \Big(exp\Big[-\frac{t-t_j}{\tau}\Big]-exp\Big[-\frac{t-t_j}{\tau_s}\Big]\Big)
    \end{equation}
where
 $\tau$ ($\tau_s$) is the decay time constant of membrane integration (synaptic currents) and $V_0$ normalizes the PSP kernels to 1, such that $\omega_j$ yields unitary PSP amplitudes. If $V(t)$ crosses the voltage threshold $V_{thr}$, the neuron spikes and the voltage is reset to $V_{rest}$. In the following we set $V_{rest} = 0$ V, $V_{0} = 1$ V, $V_{thr} = 1$ V, $\tau = 15$ ms, and $\tau_s = 3.75$ ms.  
 
In the training phase, the $V$ and $N$ tempotrons learn their corresponding task through synaptic updates, each of them performed after the presentation of the input. In the case of a correct classification, we potentiate the connections of the tempotron with the presynaptic neurons that lead to the desired firing, while the weights of the silent tempotron remain unchanged. On the other hand, in the case of an  uncorrect classification, two distinct cases may occur. If the tempotron $V$ (resp. $N$) does not respond to the presentation of the pattern $V$ (resp. $N$), then its connections with the presynaptic neurons spiking during the incoming pattern are reinforced by
    $\Delta \omega_j = \lambda \sum_{t_j < t_{max}} K(t_{max}-t_j)$,
where $\lambda$ is the learning rate and $t_{max}$ is the time at which the postsynaptic potential reaches its maximum value. If, instead, the tempotron $N$ (resp. $V$) do respond upon the presentation of the pattern $V$ (resp. $N$), we depress the synapses contributing to the erroneous firing by 
    $\Delta \omega_j = -\lambda \sum_{t_j < t_{spike}} K(t_{spike}-t_j)$,
with $t_{spike}$ being the firing time of the postsynaptic neuron. By considering only the spikes preceding $t_{max}$ or $t_{spike}$, only the connections with the presynaptic neurons which substantially contribute to the rise of $V(t)$ are updated.

Furthermore, we included the Maximum-Margin modification to reduce overfitting and improve the classification accuracy of the tempotron learning rule. In practice, a hard margin $\Delta$ is introduced during the training, thereby making the classification task more difficult to both the tempotrons and forcing them to learn additional features. The voltage threshold of the neuron that is trained to spike is increased by an amount $\Delta$, while it is decreased by the same amount in the neuron that should remain silent. In this work, we set $\Delta = 0.5V$. During the testing stage, the hard margin is suppressed and the tempotrons are expected to classify the patterns in a more accurate manner. 
Also, our model is subjected to two constraints, \emph{i.e.}\ a minimum value of the voltage ($V_{min} = -1.5$ V) and a maximum absolute values for the synaptic weights ($w_{min} = -1.5$ and $w_{max} = +1.5$). This rules out exaggerated inhibitions.

\subsection{Bin encoding}

The common encoding methods Poisson and time-to-first-spike typically map the energies of the frequency bins using only a single parameter, namely the firing rate and the time-delay, respectively  \cite{ConvSNN}. Both methods emphasize the representation of frequency bins having high short-term power, through high-firing frequency and short-time delay, respectively. This is detrimental to the encoding of single frames, in which the information on all frequency bins is relevant. 

Here, we propose the novel bin encoding scheme, in which we allow for a more complex representation of signals. This is accomplished by relating each frequency bin to a group of $N_{in}$ input neurons, instead to a single neuron as done in previous works. In particular, each neuron in the group represents a range of energy values that the associated frequency bin can take. As discussed above, the log mel filterbank bins are normalized to the interval $[0,1]$. For each frequency bin the first input neuron $n_{in} = 0$ accounts for the energy in the interval $E\in [0.0,0.1)$, the second neuron $n_{in} = 1$ accounts for $E\in [0.1,0.2)$, and so on, until the last neuron of the group $n_{in} = 9$ that represents $E\in [0.9,1.0]$. 
Upon the presentation of an input speech signal, for each frequency filter the bin in which its energy value falls is computed, and the corresponding neuron in the group is activated accordingly. For example, if the 0\textsuperscript{th} frequency filter takes the value 0.36, the corresponding bin is $E_3\in[0.3,0.4)$. Thus, the neuron $n_{in} = 3$ of group 0\textsuperscript{th} is activated. This example is illustrated as red pathway in Fig.\ \ref{fig:bin_enc}. Under low-background noise conditions, a speech-containing signal is generally encoded through the activity of neurons with high indices ($n_{in} \geq 5$), and vice-versa for those signals containing either noise or silence. This allows to enhance the differences between spike patterns belonging to different classes. 

Our bin encoding method forces each frequency band to be represented by one spike. The indices of the activated neurons contain information on the energy intensity at each given frequency. Further information is included in the timing of the spikes.
To set a temporal sequence of the spikes, we attribute a shorter time delay to the filters with larger energy. The timing of the spike for each input neuron $n_{in}$ associated with a channel having a short-term power value \textit{val} is computed as
\begin{equation}
    t_{spike} = \big[(N_{in}-1-n_{in})\times t_{interval}\big] + \textit{intensity\_diff} + \textit{offset}
\end{equation}
where $t_{interval}$ is the average-time interval between the spike time of the neurons with index $n_{in}$ and $n_{in}-1$. 
The term \textit{intensity\_diff} introduces some jitter among the neurons spiking featuring the same $n_{in}$. It is computed considering the difference in intensity between \textit{val} and the lower bound of the bin
\begin{equation}
    \textit{intensity\_diff} = \displaystyle\frac{ \big(\textit{val}\times N_{in} - n_{in}\big)\times t_{interval}}{1.5}
\end{equation}
Finally, an \textit{offset} ensures the decay of the membrane voltage after the previous input presentation. We set $t_{interval} = 7.5$ ms and \textit{offset }$ = 5$ ms.

\subsection{Post-processing}

A post-processing technique was applied to the raw predictions of the classifier. If both tempotrons spiked, then we attributed the input to the class associated with the tempotron that spiked first, as this was found to be the most confident one. If none of the tempotrons spiked, then the chosen class was the one corresponding to the tempotron that reached the highest voltage. Furthermore, to take into account contextual information, we averaged the post-processed predictions over windows composed of five frames. This approach helps avoid misclassifications of single frames in long intervals of continuous speech or no-speech activity. 
\begin{figure*}
    \centering
    \includegraphics[width=0.94\textwidth]{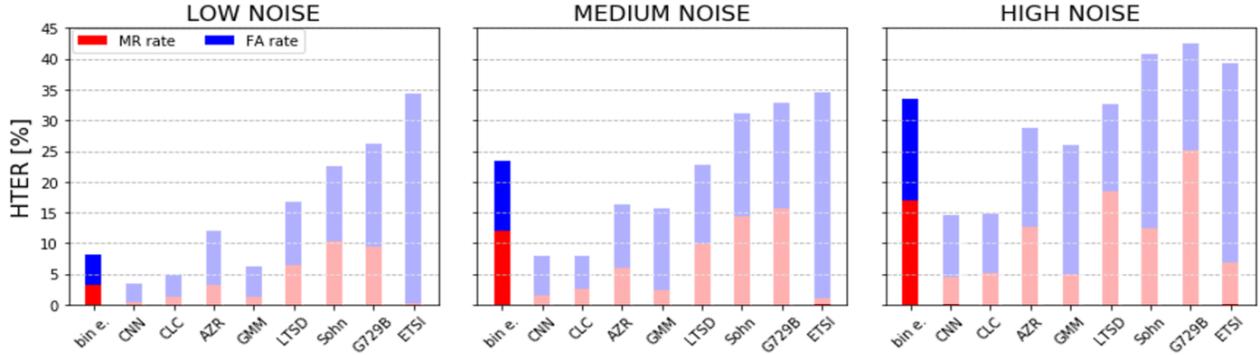}
    \caption{
    HTER performance of bin encoding VAD (bright colors) and baseline methods (shaded bars) for the three tested noise levels. The red and blue parts of the bars represent the MRs and FAs, respectively.}
    \label{fig:hter}
\end{figure*}

\subsection{Training and performance evaluation}
The synaptic weights, \emph{i.e.}\ the parameters to be learnt during training, were initially sampled from the uniform distribution $U(0,0.05)$. We presented to the network a single frame of the log mel filterbank features labeled as $V$ and $N$. The frames were randomly sampled from training data and arranged into 120 groups of 200 samples. The learning rate was initially set to $\lambda = 0.8$ and reduced by 5\% after each group. 

We evaluated the model on the QUT-NOISE augmented TIMIT data set \cite{QUT}, which mixes speech recordings from the TIMIT clean database with recordings of several noise scenarios: cafe, car, home, street, and reverberant conditions. 
In the experimental setup, we comply with the protocol for the QUT-NOISE-TIMIT database \cite{QUT}, which indicates a division of the database in three noise levels: low (SNR = +15, +10), medium (SNR = +5, 0), and high (SNR = -5, -10). Based on the noise environment, the data set is further split in Group A for training and Group B for testing. Three models are trained and tested separately, each on a different noise level.  We assessed the perfomances of our model with the Half-Total-Error Rate (HTER) metric, which computes the equally-weighted average of false alarm rate (FA) and missed detection rate (MR) as
$HTER = 0.5\times MD + 0.5\times FA$.  \\

\section{Experimental results}
\label{sec:experiments}
We evaluate the performance of our approach against eight baseline VAD systems: CNN-based VAD (CNN) \cite{Augusto}, long term spectral divergence (LTSD) \cite{ramirez}, Sohns likelihood ratio test (LRT) VAD \cite{sohn}, a GMM based approach using mel-frequency cepstral coefficient features (GMM) \cite{QUT}, ITU-T G.729 Annex B (G.729) \cite{Benyassine}, advanced front-end (AFE) ETSI \cite{li}, Complete-linkage clustering \cite{Ghaemmaghami2015} (CLC), and autocorrelation zero-crossing rate \cite{Ghaemmaghami2010} (AZR).

Our results are given in Fig.\ \ref{fig:hter}. At low-noise levels, our model achieved a detection accuracy slightly worse than that of the best performing algorithms CNN, CLC, and GMM. At medium and high-noise level conditions, our model was found comparable to LTSD, still outperforming the Sohn, ETSI and G.729 methods.
We stress that the aim of this work is not to establish state-of-the-art accuracy, rather to firstly demonstrate that SNNs can be employed to build a low-power VAD model. 
The bin encoding, unlike the Poisson encoding, presents the advantage of a constant number of spikes. Hence, the energy consumption for a time interval is independent of the nature of the signal and can therefore be precisely predicted. 
Here we provide an estimate of the power consumption of our model which takes into account only the dynamic power, and neglects the cost of running the filters and extracting the bin features.
We based our estimate on the energy measurements on the neuromorphic Intel chip Loihi ~\cite{Loihi}.
The energy consumption per second is, as in \cite{Hunsberger}: 
\begin{equation}
    E_{SNN} = \left( E_{\textit{SOP}}\times\tfrac{\textit{SOP}}{frame}+E_{\textit{n.update}}\times\tfrac{\textit{n.update}}{frame} \right)\times\tfrac{\textit{frames}}{s}
\end{equation}
In our model, No.synaptic operations (SOP) per frame are 129$\times$2, No. neuron updates per frame are 129 active and 1153 inactive, and No. frames per second is 50. 
Hence, we obtain $E_{SNN} \sim 3.8 \mu$W.
Table \ref{tab:energy} lists the energy consumption of our model along with state-of-the-art low-power VADs. We stress that these systems are already running on ASIC chips and their power estimate includes the cost of bin features generation, unlike our lower bound Loihi-based estimate.
\begin{table}[!b]
\footnotesize

\small
\centering
 \begin{tabular}{| l || l |} 
 \hline
 \textbf{Method} & \textbf{Power consumption} \\ [0.5ex] 
 \hline \hline 
 \textbf{Bin encoding} & $3.8 \mu$W (lower bound) \\
 \hline
 \textbf{Yang et al }\cite{Yang} &  $1 \mu$W\\
 \hline
 \textbf{Price et al }\cite{Price} &  $22\mu$W\\
 \hline
 \textbf{Meoni et al }\cite{Meoni} &  $559 \mu$W\\
 \hline
 \end{tabular}
 \caption{Energy consumption of bin encoding and baselines.}
 \label{tab:energy}
\end{table}

\begin{figure}
    \centering
    \includegraphics[width=0.38\textwidth]{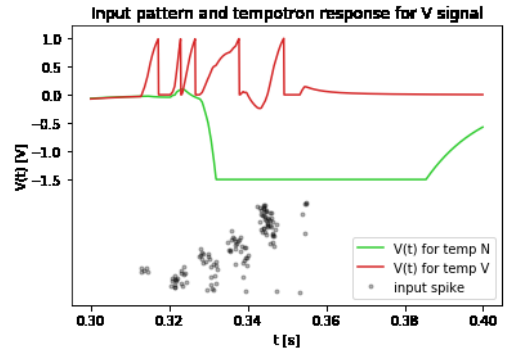}
    \caption{Early response of the $V$ tempotron upon the presentation of a voiced input.}
    \label{fig:raster}
\end{figure}

A further advantage is the possibility of training our framework only on a portion of the QUT-NOISE-TIMIT corpus. This results in both a faster training and the possibility of training the model on a new small data set.
Lastly, our approach exhibits the remarkable feature of an early decision. In fact, the tempotrons do not need to wait until the presentation of the entire input pattern is completed in order to fire. In Fig.\ \ref{fig:raster} the spike pattern encoding a voiced frame is reported along with the voltage trace of the tempotrons. The V tempotron fires after few input spikes, allowing for an early classification. 
The latency of the model is mainly reduced in the presence of speech. Indeed, the spikes at the beginning of the pattern are associated with high short-term energy bins, thus with speech features. This is seen to be especially true at high SNRs.

\section{Conclusion}
\label{ref:conclusion}
We propose a SNN architecture for VAD. We developed
a novel method for encoding the filterbank single frame features into spike patterns. We next integrated the new encoding scheme into a simple bilayer architecture which we evaluated on the QUT-NOISE-TIMIT data set. 
We showed that exploiting the power-efficiency properties of SNNs enables the design of VAD systems working at constant low power and achieving performances comparable to the state-of-the-art methods. Our framework exhibits a low latency and relies on relatively small training data.


\vfill\pagebreak

\bibliographystyle{IEEEbib}
\bibliography{strings,refs}

\begin{thebibliography}{10}

\bibitem{Blouw}
P.~Blouw, X.~Choo, E.~Hunsberger, and C.~Eliasmith,
\newblock ``Benchmarking keyword spotting efficiency on neuromorphic
  hardware,''
\newblock {\em CoRR}, vol. abs/1812.01739, 2018.

\bibitem{Loihi}
M.~Davies, N.~Srinivasa, T.~Lin, G.~Chinya, P.~Joshi, A.~Lines, Andreas W., and
  H.~Wang,
\newblock ``Loihi: A neuromorphic manycore processor with on-chip learning,''
\newblock {\em IEEE Micro}, vol. PP, pp. 1--1, 01 2018.

\bibitem{Price}
Michael Price, James Glass, and A.~P.~Chandrakasan,
\newblock ``A low-power speech recognizer and voice activity detector using
  deep neural networks,''
\newblock {\em IEEE Journal of Solid-State Circuits}, pp. 1--10, 10 2017.

\bibitem{Meoni}
G.~{Meoni}, L.~{Pilato}, and L.~{Fanucci},
\newblock ``A low power voice activity detector for portable applications,''
\newblock in {\em 2018 14th Conference on Ph.D. Research in Microelectronics
  and Electronics (PRIME)}, July 2018, pp. 41--44.

\bibitem{Yang}
M.~{Yang}, C.~{Yeh}, Y.~{Zhou}, J.~P. {Cerqueira}, A.~A. {Lazar}, and
  M.~{Seok},
\newblock ``A 1$\mu$w voice activity detector using analog feature extraction
  and digital deep neural network,''
\newblock in {\em 2018 IEEE International Solid - State Circuits Conference -
  (ISSCC)}, Feb 2018, pp. 346--348.

\bibitem{TavanaeiM17}
A.~Tavanaei and A.~S. Maida,
\newblock ``Bio-inspired multi-layer spiking neural network extracts
  discriminative features from speech signals,''
\newblock {\em CoRR}, vol. abs/1706.03170, 2017.

\bibitem{Yu}
Q.~Yu, Y.~Yao, L.~Wang, H.~Tang, J.~Dang, and K.~Chen Tan,
\newblock ``Robust environmental sound recognition with sparse key-point
  encoding and efficient multi-spike learning,''
\newblock {\em CoRR}, vol. abs/1902.01094, 2019.

\bibitem{Tavanei_Maida}
Amirhossein Tavanaei and Anthony~S. Maida,
\newblock ``A spiking network that learns to extract spike signatures from
  speech signals,''
\newblock {\em CoRR}, vol. abs/1606.00802, 2016.

\bibitem{Hazan}
H.~Hazan, D.~J. Saunders, D.~T. Sanghavi, H.~T. Siegelmann, and R.~Kozma,
\newblock ``Unsupervised learning with self-organizing spiking neural
  networks,''
\newblock {\em CoRR}, vol. abs/1807.09374, 2018.

\bibitem{Rumbell}
T.~Rumbell, S.~Denham, and T.~Wennekers,
\newblock ``A spiking self-organizing map combining stdp, oscillations, and
  continuous learning,''
\newblock {\em Neural Networks and Learning Systems, IEEE Transactions on},
  vol. 25, pp. 894--907, 05 2014.

\bibitem{VerstraetenLSM}
D.~Verstraeten, B.~Schrauwen, and D.~Stroobandt,
\newblock ``Isolated word recognition using a liquid state machine,''
\newblock 01 2005, pp. 435--440.

\bibitem{Verstraeten}
D.~Verstraeten, B.~Schrauwen, and D.~Stroobandt,
\newblock ``Reservoir-based techniques for speech recognition,''
\newblock {\em The 2006 IEEE International Joint Conference on Neural Network
  Proceedings}, pp. 1050--1053, 2006.

\bibitem{ConvSNN}
M.~Dong, X.~Huang, and B.~Xu,
\newblock ``Unsupervised speech recognition through spike-timing-dependent
  plasticity in a convolutional spiking neural network,''
\newblock {\em PLOS ONE}, vol. 13, no. 11, pp. 1--19, 11 2018.

\bibitem{SNNRobustSoundClassific}
J.~Wu, Y.~Chua, M.~Zhang, H.~Li, and K.~C. Tan,
\newblock ``A spiking neural network framework for robust sound
  classification,''
\newblock {\em Frontiers in Neuroscience}, vol. 12, pp. 836, 2018.

\bibitem{tempotron}
R.~Gütig and H.~Sompolinsky,
\newblock ``Gutig, r. \& sompolinsky, h. the tempotron: a neuron that learns
  spike timing-based decisions. nature neurosci. 9, 420-428,''
\newblock {\em Nature neuroscience}, vol. 9, pp. 420--8, 04 2006.

\bibitem{Cortes1995}
C.~Cortes and V.~Vapnik,
\newblock ``Support-vector networks,''
\newblock {\em Machine Learning}, vol. 20, no. 3, pp. 273--297, Sep 1995.

\bibitem{QUT}
D.~Dean, S.~Sridharan, R.~Vogt, and M.~Mason,
\newblock ``The qut-noise-timit corpus for the evaluation of voice activity
  detection algorithms,''
\newblock in {\em INTERSPEECH}, 2010.

\bibitem{Augusto}
Diego Augusto, Jose Stuchi, Ricardo Violato, and Luís Cuozzo,
\newblock {\em Exploring Convolutional Neural Networks for Voice Activity
  Detection}, pp. 37--47,
\newblock 07 2017.

\bibitem{ramirez}
J.~Ramírez, J.~Segura, C.~Benitez, Á. Torre, and A.~Rubio,
\newblock ``Efficient voice activity detection algorithms using long-term
  speech information,''
\newblock {\em Speech Communication}, vol. 42, pp. 271--287, 04 2004.

\bibitem{sohn}
{J. Sohn}, {N. S. Kim}, and {W. Sung},
\newblock ``A statistical model-based voice activity detection,''
\newblock {\em IEEE Sig. Proc. Letters}, vol. 6, no. 1, pp. 1--3, Jan 1999.

\bibitem{Benyassine}
A.~{Benyassine}, E.~{Shlomot}, H.~. {Su}, D.~{Massaloux}, C.~{Lamblin}, and
  J.~. {Petit},
\newblock ``Itu-t recommendation g.729 annex b: a silence compression scheme
  for use with g.729 optimized for v.70 digital simultaneous voice and data
  applications,''
\newblock {\em IEEE Communications Magazine}, vol. 35, no. 9, pp. 64--73, Sep.
  1997.

\bibitem{li}
J.~Li,
\newblock ``A complexity reduction of etsi advanced front-end for dsr,''
\newblock in {\em Proc. ICASSP}, January 2004.

\bibitem{Ghaemmaghami2015}
H.~Ghaemmaghami, D.~Dean, S.~Kalantari, S.~Sridharan, and C.~Fookes,
\newblock ``Complete-linkage clustering for voice activity detection in audio
  and visual speech,''
\newblock Dresden, Germany, September 2015, Interspeech 2015.

\bibitem{Ghaemmaghami2010}
H.~Ghaemmaghami, B.~Baker, R.~Vogt, and S.~Sridharan,
\newblock ``Noise robust voice activity detection using features extracted from
  the time-domain autocorrelation function,''
\newblock 01 2010, pp. 3118--3121.

\bibitem{Hunsberger}
E.~Hunsberger and C.~Eliasmith,
\newblock ``Training spiking deep networks for neuromorphic hardware,''
\newblock 11 2016.

\end{thebibliography}

\end{document}